\documentclass{iau} 
\usepackage{graphicx}

\title[Galactic Surveys in the Gaia Era]{Galactic Surveys in the Gaia Era}

\author[Rosemary F.G.~Wyse]   
{Rosemary F.G.~Wyse}

\affiliation{Johns Hopkins University, Department of Physics and Astronomy, \\ Baltimore, MD 21210, USA  \\ email: {\tt wyse@jhu.edu}}

\pubyear{2017}
\volume{330}  
\jname{Astrometry and Astrophysics in the Gaia Sky}
\editors{A.~Recio-Blanco, P.~de Laverny, A.~Brown \& T.~Prusti, eds.}
\begin{document}

\maketitle

\begin{abstract}
The final astrometric data from the Gaia mission will transform our view of the stellar content of the Galaxy, particularly when complemented with spectroscopic surveys providing stellar parameters, line-of-sight kinematics and elemental abundances. Analyses with Gaia DR1 are already demonstrating the insight gained and the promise of what is to come with future Gaia releases. I present a brief overview of results and puzzles from recent Galactic Archaeology surveys for context, focusing on the Galactic discs.
\end{abstract}

\firstsection 
\section{The Fossil Record: Galactic Archaeology}

Galactic Archaeology, or Near-Field Cosmology, is possible due to the
long life-times of low-mass stars and the fact that the kinematics and
chemical abundances of such stars contain information about conditions
at their birth. Nearby stars older than 10Gyr probe redshifts greater
than 2 and are found throughout the Galaxy and its retinue of
satellites. Photospheric elemental abundances are largely unchanged
throughout the lifetime of a star (setting aside mass transfer in
close binaries, and the very low level of pollution through accretion
of ambient material through which the star passes) and hence reflect
those of the interstellar gas from which the star formed. Certain
orbital properties, such as energy and angular momentum, are
approximate adiabatic invariants in realistic potentials and structure
in chemical and kinematic phase space persists after coordinate space
structure is erased by mixing (due to the finite velocity dispersion of the structure).  The multivariate distribution
functions of different Galactic components overlap, so that large
samples of stars with well-understood selection functions are required.  The
study of resolved stars of all ages within one galaxy (limited at the
present to members of the Local Group) as a means to decipher how
galaxies evolve is complementary to the direct study of the integrated
light of galaxies at high redshift: evolution of a few galaxies
compared to snapshots of different galaxies at different times. 

The detailed study of stellar populations in the Milky Way and
satellite galaxies (also M~31) is of particular importance to efforts
to identify signatures of dark sector physics.  Different candidate
dark matter particles make very different predictions for structure
formation on, and below, the scales of large galaxies, while on large
scalles, where gravity dominates the physics, there is little
divergence (see the review by \cite[Ostriker \& Steinhardt,
2003]{OS03}). These small scales are just those on which tensions are
found between the predictions of $\Lambda$CDM models without
fine-tuning and observations (e.g. \cite[Weinberg et
al.~2015]{weinberg}). Much current effort is directed both at
investigations of alternative dark matter candidates and
investigations of more complex/improved baryonic physics, 
particularly stellar feedback, to modify the distribution of dark matter. Baryonic physics is
imprinted on the fossil stellar populations and more sophisticated
models within the $\Lambda$CDM framework are being developed to test
against the increasingly detailed and comprehensive observational
data.

\begin{figure}[h]
\vskip 1.2 truecm
 \includegraphics[width=4in]{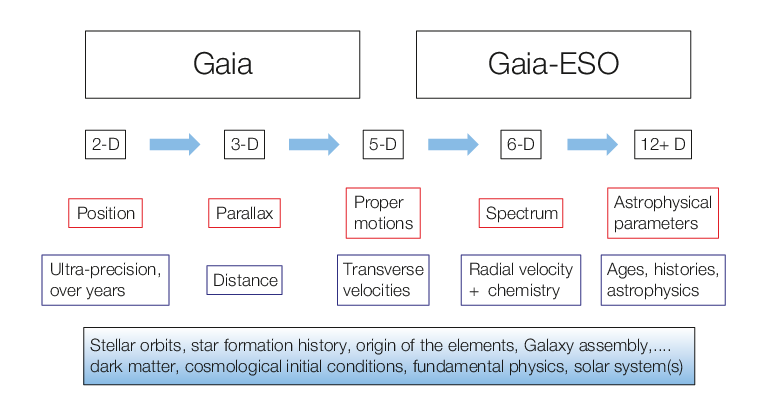} 
\vskip 0.5truecm
 \caption{Illustration of the increase in dimensionality of the parameter space within which resolved stellar populations may be studied with the combination of  astrometric and spectroscopic data. The observational quantities obtained from each type of data  are in the boxes outlined in red, with the derived astrophysical quantities in the blue-outlined boxes. The 12+ dimensions of `Astrophysical parameters' refers to the spectroscopic gravity, effective temperature and many individual elemental abundances that may be estimated from the spectra. The `big-picture' physics questions  that may then be addressed are     in the blue-shaded box below.  Adapted from Gilmore et al.~(2012), their Fig.~2.}
   \label{fig1}
\end{figure}

As I will discuss below, using the Galactic discs as examples, the distribution of stars in $n-$dimensional space is a complex function of how the Galaxy formed and evolved. 
The power of the combination of Gaia astrometric data with spectroscopic data is illustrated in
Fig.\,\ref{fig1}, modified from the overview of the Gaia-ESO survey (\cite[Gilmore et 
al., 2012]{GES}).

\begin{figure}[h]
\vskip -10.75truecm
\hskip 2.7 truein
\includegraphics[width=1.8in]{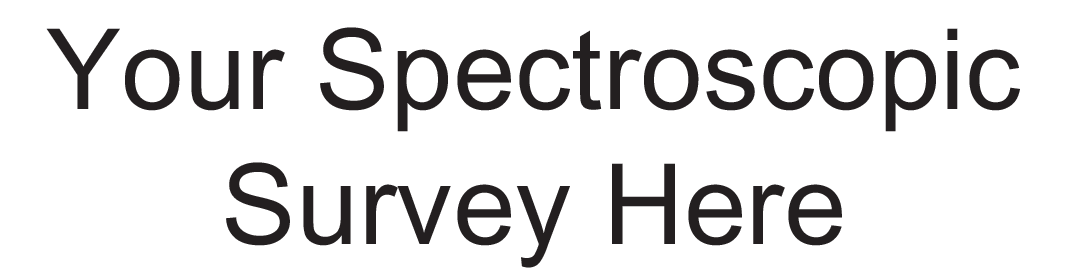}
\end{figure}

\vskip 9.5 truecm

There are several recent, on-going and planned stellar surveys using
highly multiplexed spectrographs, each with its own niche defined by
the selection function and characteristics of the spectra (such as
resolution, wavelength coverage). One hopes that consistent
conclusions will be reached from analyses of the different datasets.
Those targeting brighter stars, such as the RAVE survey (see
\cite[Kunder et al., 2017]{DR5} plus Andrea Kunder's contribution to
this volume), the GALAH survey (see \cite[Martell et
al., 2017]{GALAH}), the APOGEE survey (see Allende-Prieto's contribution to this volume) and the LAMOST survey (see Martin Smith's
contribution to this volume) have significant overlap with the sample
of the Tycho Gaia Astrometric Solution of Gaia DR1 and the range of talks and
posters in this conference give a flavour of the wealth of science
questions that can be addressed with these data, and point to the
future analyses that will be enabled with the full Gaia dataset.

On-going and planned ground-based multi-band imaging surveys  such as PanSTARRS1 (\cite[Kaiser et al.~2010]{PS1}), the Dark Energy Survey (\cite[Abbott et al.~2016]{DES}) and LSST (\cite[Abell et al.~2012]{LSST}), together with space-based  surveys from current and future missions, further extend the dimensionality of the dataset. Of course, the Gaia photometry and positions by themselves contain new information (for example see Deason's contribution to this volume for a clever use of photometry flags in Gaia DR1 to identify variable stars).  This is truly an exciting time for Galactic astrophysics.

\section{The Milky Way Discs: What can we learn from (old) disc stars? }

Thin stellar discs are fragile and can be disturbed by external
influences such as companion galaxies and mergers, in addition to
internal gravitational perturbations such as spiral arms, bars and
Giant Molecular Clouds.  Stellar systems are collisionless and thus
cannot `cool' once heated, unlike gas whuch can radiate away energy in
excited internal degrees of freedom. The vertical structure of the
thin disc - with `structure' intended to refer to all aspects,
including number density, age distribution, chemical abundance
distribution and kinematics - encodes the history of heating and minor
merging/satellite accretion, relative rates of dissipational settling
and star formation, adiabatic compression and further heating.  The
radial structure reflects the relative rates of star formation and gas
flows, as a function of location, and contains imprints of the angular
momentum distribution and re-arrangement. The thick disc probes the
earliest phase of disc star formation and the nature, and rate, of the
transition between the thick and thin discs constrain the duration of
more significant mergers and associated heating and/or turbulent
conditions, subsequent gas cooling and accretion to (re-)form a thin
disc.  Interactions with gravitational perturbations, particularly
satellites, do not only heat and thicken stellar discs, they can also
excite warps and `bending' and `breathing' modes in the thin disc (for
example \cite[Widrow et al., 2014]{wid2014}). Some radial spreading of
the disc is created as angular momentum is transferred from a
satellite to disc stars; this change in mean orbital radius occurs
together with transfer of random energy into the orbit i.e. associated
heating (e.g. \cite[Bird et al.~2012]{bird12}). Induced radial
displacement of stars can also occur without a change in the orbital
circularity, due to interactions between stars and a perturbation (such as a transient spiral arm) at the corotation resonance of the perturbation 
(\cite[Sellwood \& Binney 2002]{JSB02}); this maintains the thinness
of discs while stars move within them, and is commonly referred to as
Radial Migration (even though other types of interaction can also
cause a star's mean orbital radius to change). The outer regions of
thin discs can be built-up by outward radial migration, and the
quantification of this contribution to the stellar population at large
radius, compared to \textit{in situ\/} star formation, is of obvious
importance.

\subsection{Vertical Structure: The Local Milky Way Thick Disc}

\begin{figure}[h]
\vskip 0.2 truecm
\hskip 2.5 truecm
 \includegraphics[width=2.75in]{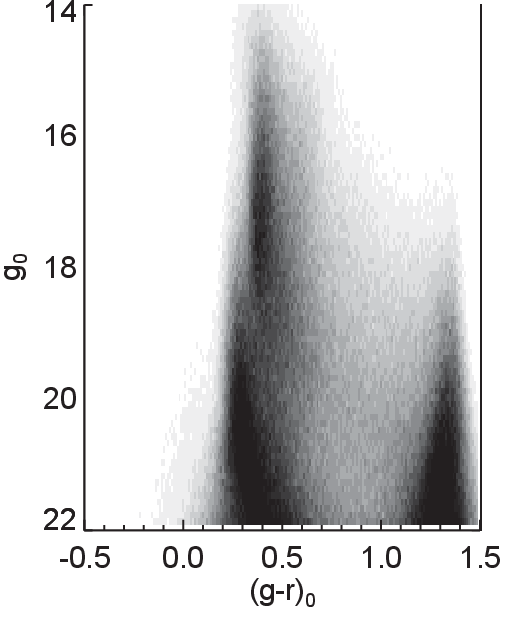} 
 \caption{Taken from \cite[Jayaraman et al. (2013)]{apoorva}, showing star counts at intermediate latitudes across the SDSS Equatorial stripe. Each of the halo and thick disc show a well-defined - and distinct - blue edge, marking the main sequence turnoffs.}
   \label{fig2}
\end{figure}

The thick disc was initially defined geometrically, through fits to
star counts at the South Galactic Pole; two separate exponentially
declining density laws were required (\cite[Gilmore \& Reid
1983]{GR83}), with the thick disc having a significantly larger scale
height than the thin disc. This geometric thick disc was subsequently
shown (by many authors, see reviews by \cite[Gilmore, Wyse \& Kuijken,
1989]{GWK}; \cite[Majewski, 1993]{maj}) to have kinematics
intermediate between those of the thin disc and stellar halo: the mean
orbital rotation velocity about the Galactic centre lagging that of
the thin disc by $\sim 50$~km/s, and the vertical velocity dispersion
being $\sim 40$~km/s, consistent with the estimated scale height of
$\sim 1$~kpc (there is a degeneracy between scale-height and local
normalization in the fits to the star counts, but the scale-height has
to be consistent with the kinematics and inferred mass surface
density).  These kinematics are too hot to have resulted from heating
due to present-day internal disc perturbations (e.g.~spiral arms,
GMCs). A discontinuous trend in age-velocity dispersion from thin to
thick disc suggested an exceptional heating event to form the thick
disc. Early (multi-object) spectroscopic studies determined a mean
metallicity for the thick disc of $\sim -0.5$~dex, with
`alpha-enhanced' elemental abundances ([$\alpha$/Fe] $> 0$). The
redder turn-off colour compared to the stellar halo (see
Fig.\,\ref{fig2}, with this metallicity, implies that most thick disc
stars are similar in age to the stellar halo, $\sim 10-12$~Gyr, thus
forming at redshifts greater than 2. Since there are stars of
significantly younger ages in the local thin disc, which would be
scattered into the thick disc should there have been a significant
merger, or indeed any strong heating event, subsequent to their
formation, the dominant old age in the (geometric) thick disc limits
any such event to have occurred only at early epochs. The derived stellar mass
of the thick disc is a significant fraction, in the range of 20\% to
50\%, of the thin disc mass, i.e. greater than $10^{10}$M$_\odot$. The narrow range of ages of the bulk of stars in the thick
disc implies a high past star-formation rate.

\subsection{Chemically Defined Thick and Thin Discs}

More precise elemental abundance measurements for larger samples of
stars have revealed that local (thin and thick) disc stars separate
into two sequences in the plane of [$\alpha$/Fe] against
[Fe/H]. Differences in the patterns of elemental abundances largely
reflects differences in star formation history (e.g. \cite[Gilmore \&
Wyse 1991]{GW91}), and the two sequences indeed are in line with the
inferred star formation histories of the thick and thin discs, albeit
that the extent of the overlap in iron abundance is large.  Further,
the mean properties of chemically defined `high-alpha' and `low-alpha'
discs compare well with those of the geometrically defined discs: the
`high-alpha' sequence is on average more metal-poor, consists of older
stars, has `hot' kinematics, and is taken to represent the thick disc,
while the `low-alpha' sequence is more metal-rich, contains young to
old stars, has `cold' kinematics and is taken to represent the thin
disc (e.g. \cite[Bensby et al. 2014]{bensby}; \cite[Martig et
al.~2016a]{martig}). Indeed, the more metal-poor stars in the
`high-alpha' sequence (which actually have enhanced values of
[$\alpha$/Fe]) have old ages and heights above the plane $\sim 1$~kpc
(\cite[Martig et al.~2016b]{martigb}) just as do stars in the
geometrically defined thick disc. The two sequences merge at the
metal-rich end(s), and there is real ambiguity about the assignment -
and origins of - the metal-rich high-alpha stars, which also tend to
be younger (e.g.~\cite[Haywood et al.~2013]{haywood13},
\cite[Chiappini et al.~2015]{chiappini15}). The younger stars at
higher heights in the outer disc (\cite[Martig et al.~2016b]{martigb}) probably reflect flaring of the thin disc. 

\begin{figure}[h]
\hskip 2.95 truecm
 \includegraphics[width=2.5in]{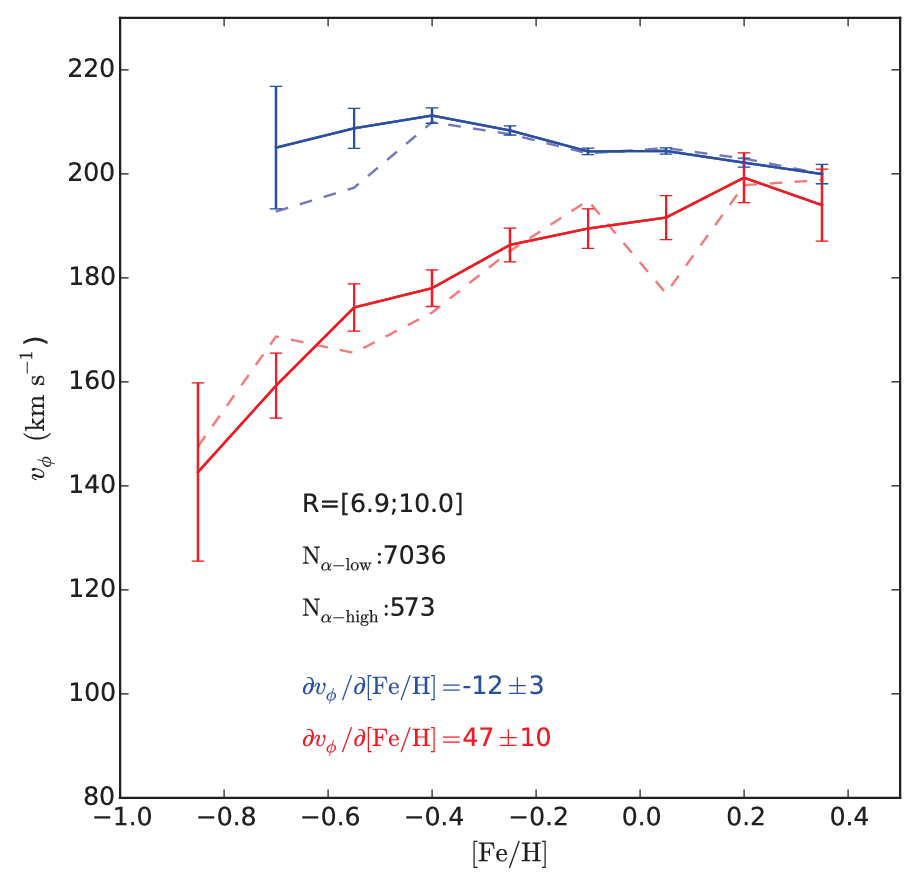} 
 \caption{Taken from \cite[Kordopatis et al.~(2017)]{APOGEE}, showing the different trends of azimuthal streaming (rotation) velocity ($v_\phi$) with iron abundance for stars in each of the two elemental abundance sequences (blue  indicates  the `low-alpha' or thin disc sequence, while red indicates the `high-alpha' or thick disc sequence). The data are from the APOGEE survey, in lines-of-sight selected to probe this velocity component without the use of tangential velocities, which are more uncertain prior to  the improved proper motions and distances that Gaia will provide.} 
   \label{fig3}
\end{figure}

The distinctiveness of the kinematics of the two elemental abundance
sequences is seen clearly in the very different trends of rotational
velocity with iron abundance, illustrated in Fig.\,\ref{fig3}. Again
we see the sequences merging at high iron abundances. The trend of
increasing rotational velocity for decreasing iron abundance in the
`low-alpha' (thin disc) sequence is understandable as a consequence of
the combination of a negative radial metallicity gradient and epicyclic motions, conserving orbital angular
momentum - a star observed
close to its perGalacticon, on an inward epicyclic excursion, will
have a higher orbital rotation velocity and have a lower metallicity
than the disc where it is observed, and \textit{vice versa\/}
for a star observed close to its maximum outward epicyclic excursion.  The
opposite trend for the `high-alpha' (thick disc) sequence must reflect
how the thick disc formed and evolved (see e.g.~\cite[Sch{\"o}nrich \&
McMillan, 2017]{SMcM}).

\subsection{The (Geometric) Thick Disc and the Earliest Phase of Disc Evolution}

Stars in the (geometric) thick disc are old, and formed at lookback
time of $\sim 10$~Gyr, corresponding to redshift $\sim 2$: the epoch
of peak cosmic star formation. The high stellar velocity dispersion
and inferred short duration of star formation implies that the thick
disc likely formed prior to equilibrium/virialization of the Milky
Way's dark halo, during the epoch of active assembly/mergers, leading
to highly turbulent conditions in the ISM (cf. \cite[Jones \& Wyse
1983]{BJRW}, \cite[Gilmore 1984]{GG84}, \cite[Brook et
al. 2012]{brook}, \cite[Bird et al.~2013]{bird13}, \cite[Ma et
al.~2017]{ma}). The derived stellar mass ($M_* > 10^{10}M_\odot$) and star-formation rate (several $M_\odot$/yr) are  similar to those derived for star-forming discs observed at redshift $\sim  2$. These high-redshift galaxies show  clumpy, turbulent ionized gas discs, in organized rotational motion with amplitude of $\sim 100-200$~km/s,  and a high internal velocity dispersion, $\sim  50-100$~km/s (e.g.~\cite[Wisnioski et al.~2015]{emily}). Lower redshift star-forming gas discs show higher values of rotational velocity to random motions, and higher specific angular momentum (\cite[Swinbank et al.~2017]{swin}). How this translates into the evolution of an individual disc consisting of stars and gas is model-dependent. The fossil record from discs in nearby galaxies provides the best guide.

\subsubsection{The Old Age of Thick Disc Stars Limits Recent Merger/Heating Events}

Mergers heat the thin stellar disc and input stars formed up to that
epoch into the (geometric) thick disc (and perhaps the stellar
halo). There are stars of a broad range of ages in the thin disc,
reflecting continuous star formation since early times. The dominant
old age of stars in the (geometric) thick disc  implies that there has been no
significant merger since the redshift at which the look-back time
equals this old age of thick disc stars. An age of $\sim 10$~Gyr means
a quiescent merger history since redshift $\sim 2$ (\cite[Wyse
2001]{rfgw}).  Such a quiet merger history is consistent with there
being no evidence of a significant dark or stellar accreted disc
(e.g. \cite[Ruchti et al., 2015]{greg}).  Minor mergers/interactions,
for example that ongoing with the Sagittarius dwarf spheroidal, affect
primarily the outer thin disc, inducing outer spiral
structure, warping and/or flaring (e.g.~\cite[Purcell et
al.~2011]{pur}).

The underlying assumption
 of this dating technique gains support  from the recent analysis by \cite[Ma
et al.~(2017)]{ma} of the age structure of the thin/thick discs formed
in their simulation of the formation of a Milky Way-mass disc galaxy
in $\Lambda$CDM. Their model galaxy incurred its last significant
merger at redshift $\sim 0.7$, a lookback time of $\sim 6$Gyr. Fig.~4 of Ma et al.~shows that across the radial
extent of the disc the mean age of stars at heights 1-2~kpc
above the mid-plane is $\sim 4-7$~Gyr, with mean metallicity in the range $-0.3$~dex to the solar
value. This mean age, for what would be identified as the geometric thick disc, indeed reflects the timing of the last significant merger  $\sim 6$Gyr ago (plus time for the merger/heating to complete), and the high mean enrichment is also consistent with a later heating event in the simulation than implied for the Milky Way.

Accurate and precise (old) stellar ages are crucial in this estimation
of the timing of the last significant merger and the parallax data
from Gaia will be extrenely important (see \cite[Tayar et
al.~2017]{2017ApJ...840...17T} for a  discussion of the biases
that must be dealt with in the estimation of ages for red giant stars 
based on astrosiesmology-based gravities). Distances are also of course fundamental to the determination of the vertical stucture. See \cite[Mackereth et
al. 2017]{2017arXiv170600018M} for a recent pre-Gaia analysis of the  age-metallicity structure of the discs from APOGEE
data. The quiescent merger history inferred for the Milky Way is not typical in $\Lambda$CDM and we also need more detailed predictions of the merger history of
typical Milky Way-mass disc galaxies, as a function of the orbit and
mass ratio of the satellite/subhalo, from galaxy formation models, to
interpret the derived observational limits.

\section{Radial Structure: the Outer Disc}

\subsection{The Ringing Disc}

A wealth of structure in the outer stellar disc has been revealed by
the imaging data of SDSS (e.g.~\cite[Belokurov et al., 2006)]{FoS})
and Pan-STARRS (e.g. \cite[Slater er al., 2014]{ps1} and \cite[Bernard
et al., 2016]{ps2}). Systematic variations in star counts above and
below the nominal mid-plane were identified by \cite[Xu et
al.~(2015)]{rings} and interpreted as rings and radial waves in the
disc. The `Monoceros Ring', which had been speculated to be a remmant
of an accreted satellite (e.g. \cite[Pe{\~n}arrubia et al.,
2005]{mon}) is clearly more simply an apparent overdensity due to
structure within the thin disc. Oscillatory kinematic features in thin
disc stars have also been identified in several kinematic surveys
(e.g.~\cite[Widrow et al., 2012]{sdss} and \cite[Williams et al.,
2013]{RAVE}). These plausibly reflect breathing/bending modes of the
perturbed thin disc (\cite[Widrow \& Bonner 2015]{bb}), excited due to
interaction by either internal perturbations, such as spiral structure
and/or the bar (\cite[Debattista, 2014]{2014MNRAS.443L...1D}) or external perturbations, perhaps the Sagittarius dwarf spheroidal (\cite[G{\'o}mez et al.,
2013]{gom}; see also \cite[Widrow et al., 2014]{wid}). These oscillations, combined with flaring of the thin disc, may extend to high enough distances from the plane that  even apparent `halo' substructure may actually consist of the perturbed thin disc (\cite[Li et al., 2017]{2017arXiv170305384L}).

There are several mechanisms by which a sub-population of thin disc
stars can be scattered/heated into the halo, such as through binary
interactions. This could be the explanation behind the discovery of rare
metal-rich high-velocity stars (e.g.~\cite[Hawkins et al.,
2015]{kh}). As is always the case for the analysis of objects with
extreme values of the parameters characterizing their parent
population (kinematics and chemistry in this case), the accuracy and
precision of those values are critical.

 The main lesson is that overdensities in star counts and kinematic
substructure are not necessarily tidal debris from satellites. Again,
a more comprehensive understanding will be available with improved
distances, three-dimensional velocities and ages from the Gaia data.

\subsection{Internal, Secular Evolution to Re-arrange Discs?}

As noted above, radial migration (Sellwood \& Binney, 
2002) can move thin-disc stars across distances that are of
order the disc scale-length, during the lifetime of the disc, without
associated kinematic heating (maintaining orbital circularity and the
thinness of the disc). This mechanism is more effective for stars on
closer-to-circular orbits, less so for populations of higher velocity
dispersion/lower angular momentum orbits (e.g. \cite[Solway, Sellwood
\& Sch\"onrich, 2012]{sss}; \cite[Vera-Ciro et al., 2014]{vera};
\cite[Daniel \& Wyse, 2017]{kated2}). The efficiency of radial
migration also obviously depends on the parameters of the
perturbation(s) driving it, such as amplitude, duty cycle, pattern
speed and wave number (e.g.~\cite[Daniel \& Wyse, 2015]{kd1};
\cite[Debattista, Roskar \& Loebman, 2017]{victor}). The chemical evolution of the disc can be strongly affected (\cite[Sch{\"o}nrich \& Binney, 2009]{2009MNRAS.396..203S}).  The existence of
a significant population of thin-disc stars in the solar neighbourhood
with super-solar metallicities is consistent with outward radial
migration bringing stars from the inner, higher-metallicity, regions
of the disc to the outer disc (e.g. \cite[Kordopatis et al.,
2015]{gkb}). The global importance of radial migration in the
evolution of discs is uncertain, with simulations showing both a
minimal influence (Bird et al., 2013), and a very
important role (\cite[Minchev et al., 2012]{2012A&A...548A.126M}). It
is clearly important to isolate the important phyical effects causing
these different conclusions, and test the predictions of the models in
detail.

\section{Concluding remarks}

 	`More data are needed': accurate and precise positions,
distances, space motions, ages, and chemical abundances are all critical
to the characterisation of the present stellar populations of the Galaxy and to
the eventual understanding of  the roles of different physical processes in its
evolution. Happily this is what the combination of astrometric data
from Gaia and data from large spectroscopic surveys promises to
deliver.  When analysing these data, we need to be clear to what we
are referring when using the terms halo/bulge/thin disc/thick disc,
since entities defined through different parameters (e.g. spatial
distribution, chemistry) may have different histories. 

 There are truly exciting times ahead: wonderful observational data for stars  plus improved simulations of galaxy formation in cosmological context(s), complemented by increasingly detailed data for high-redshift discs in formation.

\medskip

\textit{Acknowledgements:} I thank the organisers for inviting me, and the American Astronomical Society for the award of an NSF-funded International Travel Grant.


\begin{thebibliography}{}

\bibitem[]{DES}{Abbott, T. et al. (Dark Energy Survey Team)} 2016, \textit{MNRAS}, 460, 1270

\bibitem[]{LSST}{Abell, P.A. et al. (LSST)} 2012, arXiv:0912.0201


\bibitem[Bensby et al.(2014)]{bensby} {Bensby, T., Feltzing, S., \& Oey, M.~S.} 2014, \textit{A \& A}, 562, A71 

\bibitem[Bernard et al.(2016)]{ps2} {Bernard, E.~J., Ferguson, A.~M.~N., Schlafly, E.~F., et al.} 2016, \textit{mnras}, 463, 1759 

\bibitem[Bird et al.(2012)]{bird}{Bird, J.~C., Kazantzidis, S., \& Weinberg, D.~H.} 2012, \textit{MNRAS}, 420, 913 

\bibitem[Bird et al. (2013)]{bird13}{Bird, J.~C., Kazantzidis, S., Weinberg, D.~H., et al.} 2013, \textit{ApJ}, 773, 43 

\bibitem[Brook et al.(2012)]{brook}{Brook, C.~B., Stinson, G.~S., Gibson, B.~K., et al.} 2012, \textit{MNRAS}, 426, 690 



\bibitem[Chiappini et al.(2015)]{chiappini15}{Chiappini, C., Anders, F., Rodrigues, T.~S., et al.} 2015, \textit{A \& A}, 576, L12 

\bibitem[Daniel \& Wyse(2015)]{kd1} {Daniel, K.~J., \& Wyse, R.~F.~G.} 2015, \textit{MNRAS}, 447, 3576 



\bibitem[]{kd2}{Daniel, K.J. \& Wyse, R.F.G.} 2017, \textit{MNRAS}, submitted

\bibitem[Debattista(2014)]{2014MNRAS.443L...1D}{Debattista, V.~P.} 2014, \textit{MNRAS}, 443, L1 

\bibitem[Debattista et al.(2017)]{victor}{Debattista, V.~P., Roskar,
R., \& Loebman, S.~R.} 2017, in: Eds. J. H. Knapen, J. C. Lee and
A. Gil de Paz, (eds.)  `Outskirts of Galaxies', \textit{ASSL},
(Berlin: Springer), in press (arXiv:1706.01996)



\bibitem[]{GW91}{Gilmore, G., \& Wyse, R.~F.~G.} 1991, \textit{ApJL}, 367, L55 

\bibitem[Gilmore et al.(1989)]{GWK}{Gilmore, G., Wyse, R.~F.~G., \& Kuijken, K.} 1989, \textit{ARAA}, 27, 555 


\bibitem[]{GES}{Gilmore, G., et al.} 2012, \textit{ESO Messenger}, 147, 25

\bibitem[G{\'o}mez et al.(2013)]{gom}{G{\'o}mez, F.~A., Minchev, I., O'Shea, B.~W., et al.} 2013, \textit{MNRAS}, 429, 159 

\bibitem[Hawkins et al., 2015]{kh} {Hawkins, K., Kordopatis, G., Gilmore, G., et al.} 2015, \textit{MNRAS}, 447, 2046 



\bibitem[Haywood et al.(2013)]{haywood13}{Haywood, M., Di Matteo, P., Lehnert, M.~D., Katz, D., \& G{\'o}mez, A.} 2013, \textit{A \& A}, 560, A109 

\bibitem[Jones \& Wyse (1983)]{BJRW}{Jones, B.~J.~T., \& Wyse, R.~F.~G.} 1983, \textit{A \& A}, 120, 165 


\bibitem[]{PS1}{Kaiser, N., et al.~(PanSTARRS} 2010, \textit{SPIE}, 7733, id 77330E

\bibitem[Kordopatis et al.(2015)]{gka}{Kordopatis, G., Wyse, R.~F.~G., Gilmore, G., et al.} 2015, \textit{A \& A}, 582, A122 


\bibitem[Kordopatis et al.(2015)]{gkb}{Kordopatis, G., Binney, J., Gilmore, G., et al.} 2015, \textit{MNRAS}, 447, 3526 



\bibitem[]{DR5}{Kunder, A., et al.} 2017, \textit{AJ}, 153, 75

\bibitem[Li et al.(2017)]{2017arXiv170305384L}{Li, T.~S., Sheffield, A.~A., Johnston, K.~V., et al.} 2017, arXiv:1703.05384 


\bibitem[Ma et al.(2017)]{ma}{Ma, X., Hopkins, P.~F., Wetzel, A.~R., et al.} 2017, \textit{MNRAS}, 467, 2430 


\bibitem[Mackereth et al.(2017)]{2017arXiv170600018M} {Mackereth, J.~T., Bovy, J., Schiavon, R.~P., et al.} 2017, arXiv:1706.00018 




\bibitem[Majewski(1993)]{maj} {Majewski, S.~R.} 1993, \textit{ARAA}, 31, 575 
\bibitem[]{GALAH}{Martell, S., et al.} 2017, \textit{MNRAS}, 465, 3203



\bibitem[]{martig} {Martig, M., Fouesneau, M., Rix, H.-W., et al.} 2016a, \textit{MNRAS}, 456, 3655 

\bibitem[]{martigb}{Martig, M., Minchev, I., Ness, M., Fouesneau, M., \& Rix, H.-W.} 2016b, \textit{ApJ}, 831, 139 

\bibitem[Minchev et al.(2012)]{2012A&A...548A.126M}{Minchev, I., Famaey, B., Quillen, A.~C., et al.} 2012, \textit{A \& A}, 548, A126 



\bibitem[Ostriker \& Steinhardt (2003)]{OS03}{Ostriker, J.P. \& Steinhardt, P.} 2003, \textit{Science}, 300, 1909

\bibitem[Pe{\~n}arrubia et al.(2005)]{mon} {Pe{\~n}arrubia, J., Mart{\'{\i}}nez-Delgado, D., Rix, H.~W., et al.} 2005, \textit{ApJ}, 626, 128 

\bibitem[Purcell et al.(2011)]{pur}{Purcell, C.~W., Bullock, J.~S., Tollerud, E., Rocha, M. \& Chakrabarti, S.} 2011, \textit{Nature}, 477, 301 

\bibitem[Sch{\"o}nrich \& Binney(2009)]{2009MNRAS.396..203S}{Sch{\"o}nrich, R., \& Binney, J.} 2009, \textit{MNRAS}, 396, 203 



\bibitem[Sch{\"o}nrich \& McMillan(2017)]{SMcM} {Sch{\"o}nrich, R., \& McMillan, P.~J.} 2017, \textit{MNRAS}, 467, 1154 

\bibitem[]{JSB02}{Sellwood, J.~A., \& Binney, J.~J.} 2002, \textit{MNRAS}, 336, 785 

\bibitem[Slater et al.(2014)]{ps1}{Slater, C.~T., Bell, E.~F., Schlafly, E.~F., et al.} 2014, \textit{ApJ}, 791, 9 

\bibitem[Solway et al.(2012)]{sss}{Solway, M., Sellwood, J.~A., \& Sch{\"o}nrich, R.} 2012, \textit{MNRAS}, 422, 1363 


\bibitem[Swinbank et al.(2017)]{swin}{Swinbank, A.~M., Harrison, C.~M., Trayford, J., et al.} 2017, \textit{MNRAS}, 467, 3140 

\bibitem[Tayar et al.(2017)]{2017ApJ...840...17T} {Tayar, J., Somers, G., Pinsonneault, M.~H., et al.} 2017, \textit{ApJ}, 840, 17 

\bibitem[Vera-Ciro et al.(2014)]{vera}{Vera-Ciro, C., D'Onghia, E., Navarro, J., \& Abadi, M.} 2014, \textit{ApJ}, 794, 173 


\bibitem[]{weinberg}{Weinberg, D.~H., Bullock, J.~S., Governato, F., et al.} 2015, \textit{PNAS}, 112, 12249 

\bibitem[Widrow et al.(2012)]{sdss}{Widrow, L.~M., Gardner, S., Yanny, B., Dodelson, S., \& Chen, H.-Y.} 2012, \textit{ApJL}, 750, L41 

\bibitem[Widrow et al.(2014)]{wid} {Widrow, L.~M., Barber, J., Chequers, M.~H., \& Cheng, E.} 2014, \textit{MNRAS}, 440, 1971 

\bibitem[Widrow \& Bonner(2015)]{bb}{Widrow, L.~M., \& Bonner, G.} 2015, \textit{MNRAS}, 450, 266 


\bibitem[Williams et al.(2013)]{RAVE}{Williams, M.~E.~K., Steinmetz, M., Binney, J., et al.} 2013, \textit{MNRAS}, 436, 101 

\bibitem[Wyse(2001)]{rfgw}{Wyse, R.~F.~G.} 2001, in: J.~G. Funes \& E.~M. Corsini (eds.) `Galaxy Discs and Disc Galaxies', \textit{Astronomical Society of the Pacific Conference Series}, 230, (San Francisco: ASP), p71

\bibitem[Xu et al.(2015)]{rings}{Xu, Y., Newberg, H.~J., Carlin, J.~L., et al.} 2015, \textit{ApJ}, 801, 105 

\end{thebibliography}
\end{document}